%% file: usenix2021_SOUPS.tex
\documentclass[letterpaper,twocolumn,10pt]{article}
\usepackage{usenix2021_SOUPS}
\usepackage[most]{tcolorbox}
\usepackage{csquotes}
\usepackage{graphics}
\usepackage{enumitem}
\setitemize{noitemsep,topsep=6pt,parsep=3pt,partopsep=3pt}

\usepackage{tikz}
\usepackage{amsmath}

\usepackage{filecontents}

\definecolor{colorbar}{HTML}{1E7CC9}
\usepackage[most]{tcolorbox}
\usepackage{csquotes}
\usepackage{graphics}
\tcbset{textmarker/.style={%
        enhanced,
        parbox=true,boxrule=0mm,boxsep=0mm,arc=0mm,
        outer arc=0mm,left=3mm,right=3mm,top=7pt,bottom=7pt,
        toptitle=1mm,bottomtitle=1mm}}
\newtcolorbox{hiBox}{textmarker,
    borderline west={6pt}{0pt}{colorbar},
    colback=colorbar!10!white}

\newcommand{\highlightbox}[1]{\begin{hiBox} #1 \end{hiBox}}
\usepackage{booktabs}

\begin{document}

\date{}

\title{\Large \bf Observations From an Online Security Competition \\and Its Implications on Crowdsourced Security}

\def\plainauthor{Author name(s) for PDF metadata. Don't forget to anonymize for submission!}

\author{
{\rm Alejandro Cuevas}\\
Carnegie Mellon University
\and
{\rm Emma Hogan}\\
University of California, San Diego
\and
{\rm Hanan Hibshi}\\
Carnegie Mellon University
\and
{\rm Nicolas Christin}\\
Carnegie Mellon University
} 

\maketitle
\thecopyright

\begin{abstract}
The crowd sourced security industry, particularly bug bounty programs, has grown dramatically over the past years and has become the main source of software security reviews for many companies. However, the academic literature has largely omitted security teams, particularly in crowd work contexts. As such, we know very little about how distributed security teams organize, collaborate, and what technology needs they have. We fill this gap by conducting focus groups with the top five teams (out of 18,201 participating teams) of a computer security Capture-the-Flag (CTF) competition. We find that these teams adopted a set of strategies centered on specialties, which allowed them to reduce issues relating to dispersion, double work, and lack of previous collaboration. Observing the current issues of a model centered on individual workers in security crowd work platforms, our study cases that scaling security work to teams is feasible and beneficial. Finally, we identify various areas which warrant future work, such as issues of social identity in high-skilled crowd work environments.
\end{abstract}

\input{1-introduction}

\input{2-related-work}
\input{3-background}

\input{4-methodology}
\input{5-results}
\input{6-discussion}

\input{7-conclusion}

\bibliographystyle{plain}
\bibliography{references}

\end{document}

%% file: 1-introduction.tex
\section{Introduction}
In the 2019 iteration of picoCTF, a two-week computer security competition with over 90,000 participants, the top three ranking teams presented a set of counter-intuitive properties. 
Members were all geographically distributed, their collaboration was fully online, and some team members did not even know each other, let alone had collaborated before. 
Yet, these teams adopted a set of strategies that allowed them to surpass thousands of other competing teams. 
While the research literature on distributed teams may provide hints on the factors that contributed to these successful collaborations~\cite{espinosa2003impact,olson2000distance, herbsleb2003empirical}, little attention has been paid to how \emph{security engineering} teams collaborate. 
In particular,
there is scant research on the specific practices of teams that join together online, without much/any prior interaction, to complete short-term security tasks.
What can we learn from the practices of these ``short-term security teams,'' particularly in the context of security crowdsourcing?

Security engineering manifests in a variety of forms across industry teams and may cover a wide variety of functions. In the context of this study, the functions that we consider are evaluation, vulnerability assessment, exploitation analysis, and digital forensics, as categorized by the NIST~\cite{newhouse2017national}. These skill sets directly overlap with those applied across many capture-the-flag competitions (CTFs), which have made CTFs a popular tool for many ends: screening potential employees (similar to coding interviews for software engineering roles)~\cite{2016facebook,armour2017announcing}, teaching in corporate, academic, and military settings~\cite{mirkovic2014class, jin2018game, adams2009collective, votipka2018toward}, and training future bug bounty hunters~\cite{hacker1012020,levelup}. For these reasons, we argue that CTF teams are a naturally good proxy to study security engineering teams, especially in distributed contexts and for short-term security tasks, such as screening code for bugs.

Today, many companies have opted to offload or complement the functions described above through crowdsourced security tasks. Bug bounty programs are among the most popular offering, although penetration testing is also increasingly gaining popularity~\cite{2019hacker-powered}. 
Companies, through bug bounty programs, invite crowds of security researchers to review their software, and issue payments to researchers who report bugs. 
Due to the cost efficiency of such programs~\cite{finifter2013empirical}, the crowd sourced security industry has been dramatically increasing over the past couple of years. 
In 2019, HackerOne reported that hacker-powered security programs increased by at least 30\% across each region of the world~\cite{2019hacker-powered}, while BugCrowd reported a 92\% increase in reported vulnerabilities~\cite{bugcrowd2019priority}. 
However, current crowdsourced security offerings focus only on harnessing work at the individual level (i.e., slicing and distributing work to individuals), which introduces issues of work replication, reduced report quality due to competition, and attrition~\cite{zhao2016crowdsourced}.

Scholars focusing on security crowd work have so far primarily focused on rigorous, quantitative modeling, mainly from an economics perspective~\cite{zhao2016crowdsourced, laszka2016banishing}, to tackle this problem. 
While economics-inspired approaches are relevant, we argue that the problem is ripe for investigation from a collaborative-work perspective as well, particularly given the prevalence of CTF teams who work together to efficiently solve similar problems to those found in BBPs.
In discussing the future of crowd work, Kittur et al. propose a framework to scale work to support more complex tasks and increase efficiency, among other gains~\cite{kittur2013future}. 
However, can we begin to conceptualize the future of security crowd work under this framework knowing next to nothing about how security teams operate and what their needs are? 
We attempt to fill this gap by focusing on security teams which embody the characteristics that teams in security crowd work would have. 

To understand the dynamics, practices, and needs of high-performing short-term security teams, we conducted focus groups with the top five ranking teams from the picoCTF 2019 competition. 
Our study is exploratory and seeks to get a wide view of the practices of short-term security teams, from their formation to their motivations, with a focus on the practices that enabled their high performance. 
Our research questions are as follows:
\begin{itemize}
    \item \textbf{RQ1:} \textit{How, why, and through which assembly mechanisms do these teams form?}
    \item \textbf{RQ2:} \textit{What technological needs do teams have and how do they address them?}
    \item \textbf{RQ3:} \textit{How do teams scope and distribute tasks, roles, and responsibilities?}
    \item \textbf{RQ4:} \textit{What motivations and limitations do teams have?}
    \item \textbf{RQ5:} \textit{What factors contribute to performance according to the teams?}
\end{itemize}

Our results offer a first view on the practices of security teams in a distributed collaboration context. 
At a high level, we note how teams organically have adopted various strategies and processes which readily fit the more advanced framework of crowd work proposed by Kittur et al.~\cite{kittur2013future}, particularly in terms of task decomposition, hierarchical/reputational organization, task assignment, and collaboration. 
For instance, we find that teams adopted a role-based approach based on specialty areas, to split and assign tasks. 
We also find that assembly and recruitment was heavily-based on role and reputation. 
Looking at the current bug-bounty program models and their limitations, our observations indicate that scaling security crowd work to teams is feasible and potentially beneficial in various aspects, both for the individuals, as well as for the work platform.
We conclude our study discussing other salient observations and identifying various areas which merit future work.

%% file: 2-related-work.tex
\section{Related Work}
\label{sec:related-work}
In this section, we review research that investigated geographically-distributed software engineering teams, security teams, and crowdsourced security. 
\subsection{Distributed Teams}
The challenges of virtual or geographically-distributed teams have been well studied and reviewed, particularly for software engineering teams~\cite{abarca2020working,gilson2015virtual,powell2004virtual}. 
Since we are not aware of prior research on security teams, we survey the literature for constructs that can guide our exploration on the factors that affect virtual teams' performance. 
Abarca et al.\ conducted a systematic literature review of 2\,354 studies on virtual teams between 2015 and 2019~\cite{abarca2020working}, following the review by Gilson et al.\ which covered work from 2004 to 2014~\cite{gilson2015virtual}, as well as the review by Powell et al.\ in 2004~\cite{powell2004virtual}. 
We describe the main constructs that resulted from these reviews.

\textit{Geographic Distribution and Communication:} Dispersion reduces the amount of communication across teams, increasing the difficulty in coordination and awareness, increasing delays~\cite{olson2000distance}. Beyond decreased communication, however, greater spatial distribution also introduces temporal challenges. Less time overlap makes it harder for teams to adjust when unexpected problems arise and makes communication more prone to breakdown~\cite{espinosa2003impact,espinosa2006effect}. In the case of software engineering teams, distribution leads to more software failures~\cite{herbsleb2003empirical}. Herbsleb and Mockus posit that increased awareness, communication, and better work distribution can reduce cross-site delays~\cite{herbsleb2003empirical}. Newer studies have looked at how technologies that facilitate coordination and resource sharing can help distributed teams cope with dispersion~\cite{warshaw2016when}.

\textit{Trust:} Trust is defined as ``as an individual's willingness to become vulnerable to the actions of others with the expectations that others will follow through on their commitments~\cite{robert2016monitoring,mayer1995integrative}.'' In a study of globally distributed development teams, Al-Ani et al.\ found that participants described trust as expectations from their colleagues, such as in terms of technical competency~\cite{alani2013globally}. 
Trust is in particular cultivated when teammates fulfill the positive expectations of their team members~\cite{piccoli2004virtual,robert2009individual}. Increased trust fosters knowledge sharing and coordination, and leads to greater performance~\cite{jones1998experience,mayer2005trust,schaubroeck2011cognition}.


\textit{Types of Tasks and Interdependence:} The nature of tasks and their assignment impacts team performance. When members work on tasks with more interdependence, more communication and coordination is needed, particularly if these are unplanned. 
In software engineering teams, the high amount of interdependence between tasks is among the variables that most influence team performance~\cite{herbsleb2003empirical,langfred2005autonomy}.

\textit{Cohesion and Familiarity:} Cohesion relates to the sense of unity in a team~\cite{abarca2020working,chidambaram1996relational}. 
Physical distance can make it harder for teammates to get to know (i.e., increase familiarity with) each other~\cite{salisbury2006cohesion}, but given enough time distributed teams may still become cohesive through online communications~\cite{chidambaram1996relational}. 
Past studies have found that cohesion leads to better performance~\cite{lurey2001empirical}. 
Similarly, knowing of how other members work or who knows (``team awareness'') can also increase the team's performance~\cite{dourish1992awareness}. 
As such, prior collaboration, such as ``team dating''(i.e., interacting in brief tasks before choosing teams) can help members do better, particularly in ad hoc scenarios~\cite{lykourentzou2017team}.

\textit{Leadership and Motivations:} Team leadership can be an effective tool for increasing motivation and coordination in a team~\cite{bradford2002typology}. 
Past research, however, has found that the lack of face-to-face contact attenuates the positive effects of hierarchical leadership~\cite{bradford2002typology,purvanova2009transformational}. 
Some scholars suggest that structural supports (such as fair and reliable reward systems) and shared leadership can complement hierarchical leadership in virtual settings to increase performance~\cite{bradford2002typology,hinds2002distributed}. 
Particularly, the performance gain of sharing leadership across members is based on the premise that it empowers individual members~\cite{kirkman2004impact}. Hoch and Kozlowski found empirical support for the performance advantages of sharing leadership~\cite{hoch2014leading}, while Zhu et al. found evidence that leadership behavior by all members in online communities increased members' motivations~\cite{zhu2012effectiveness}.

\subsection{Security Teams}
In a 2017 NIST special publication on creating a cybersecurity workforce framework, computer security work was split in 7 categories: securely provision (SP), operate and maintain (OM), oversee and govern (OV), protect and defend (PR), analyze (AN), collect and operate (CO), investigate (IN)~\cite{newhouse2017national}. 
Each of these categories is then split into specialty areas~\cite{newhouse2017national}. 
For the purposes of this paper, the specialty areas we consider when discussing security teams in the context of CTFs and bug-bounty programs are the following: test and evaluation (TST, part of SP), vulnerability assessment and management (VAM, part of PR), exploitation analysis (EXP, part of AN), and digital forensics (FOR, part of IN).

Most of the academic work on security teams has focused on computer security incident response teams (CSIRTs). CSIRTs are part of the PR category, and the incident response (CIR) specialty area. Work CSIRTs has focused on: exploring it from an organizational psychology perspective~\cite{chen2014organizational}, formalizing management capabilities~\cite{ruefle2014computer}, and strategies to develop shared knowledge and increase performance~\cite{steinke2015improving}. Beyond CSIRTs, Henshel et al. identify the need for and propose an assessment model for quantifying cyber defense teams' proficiency~\cite{henshel2016predicting}. Kokulu et al. studied Security Operation Centers (teams of security analysts who monitor, prevent, report, and respond to security attacks) and found that there are disagreements between managers and analysts which threaten the efficiency and effectiveness of these security teams~\cite{kokulu2019matched}.

\subsection{Bug Bounties and Crowdsourced Security}

Bug bounty programs are programs offered by organizations through which external security professionals are given the opportunity to look for security bugs across the organizations' software. 
Security professionals who find a bug can then submit a report to the organization and depending on the criticality of the bug, receive commensurate compensation (i.e., a bounty). 
The idea is that harnessing the wisdom of the crowd is more economically efficient and also results in more discovered bugs. Both claims have been validated empirically, the former by Finifter et al~\cite{finifter2013empirical} and Walshe and Simpson~\cite{walshe2020empirical}. and the latter by Maillart et al.~\cite{maillart2017given}. 

Bug bounties are possibly the most popular crowdsourced security offering, with over 1400 participating organizations, 450,000 registered hackers, and more than US\$62M paid in bounties in the HackerOne platform alone~\cite{2019hacker-powered}. 
However, 
BBPs suffer from a high number of duplicate submissions (30-40\%) due to the competition between researchers. 
Researchers may devote weeks to the same security bug only for the first submission to be rewarded~\cite{zhao2016crowdsourced}. 
This has caused researchers to quickly flock to new programs, prioritizing higher payouts and easier bugs~\cite{maillart2017given}.

%% file: 3-background.tex
\section{Background}
We provide an overview of the nature and motivation of capture-the-flag competitions, as well as their current usage for recruitment and education. Following, we describe the picoCTF competition and its participants. We provide a sample challenge from the competition in Figure~\ref{fig:pico_challenge} and describe its solution to provide context on players' skill sets and expertise.

\subsection{Capture-the-flag Competitions}
Capture-the-flag (CTF) competitions are events where teams of computer security experts and/or students compete to showcase their skills across various areas of security, such as memory corruption, web security, and cryptography. In jeopardy-style CTFs, puzzles typically involve finding and exploiting vulnerabilities, reverse engineering binaries, or finding information through digital forensics tools. CTFs have been steadily gaining popularity, with over 170 competitions planned for 2020 and more being announced every week~\cite{2020ctfd}. While originally conceptualized as an entertaining battle of wits between talented hackers, CTFs have become much more. The NSA, along with companies like Google and Facebook, organize CTF competitions to identify and recruit security talent~\cite{2016facebook,2020nsa,armour2017announcing}. Across schools and colleges, CTF problems have been gaining popularity as a way to teach security concepts through hands-on exercises~\cite{mirkovic2014class}. And competitions are organized to foster interest in a field that is seeing a shortage in security professionals~\cite{jin2018game}. Finally, bug bounty platforms have also shifted to CTFs to provide education to bug bounty hunters~\cite{hacker1012020,levelup}. With CTFs having such a quintessential role in the education and recruitment of security professionals, we argue they are a suitable environment to study the collaboration of security teams.

\subsection{The picoCTF Competition}

picoCTF is a competition organized and run by Carnegie Mellon University (CMU). The organizers provided the research team with competition data and facilitated the recruitment of the top teams.
While the \$14,000 prize pool is restricted to middle/high school students in the US, the competition is open and free of cost to participants of all ages from around the world. 
In 2019, 46,052 registered teams, of which 18,201 had at least one successful submission.
The competition was open for 2 weeks, from September 26th to October 11th, 2019.
Teams earned points by completing challenges. In case several teams completed all challenges, their final ranking was determined by the speed at which they solved these challenges. 
The competition consisted of 124 problems, broken down across 6 categories: general, cryptography, forensics, reverse engineering, binary exploitation, and web exploitation. 
In the the Global rankings (which includes all registered teams), only 13 teams finished all challenges in the allotted time. 
While picoCTF provides beginner-friendly introductory challenges, its toughest challenges require deep technical knowledge similar to the knowledge needed to work on real systems. 
For example, Figure~\ref{fig:pico_challenge} is an example of a technical problem lifted from the competition. 
To solve this problem, participants have to be comfortable using professional reverse engineering tools, such as Ghidra~\footnote{https://ghidra-sre.org/}, 
and be familiar with the GNU C Library (\texttt{glibc})~\footnote{https://www.gnu.org/software/libc/}, 
heap management techniques (in this case, \texttt{tcache}~\cite{memory_gnu}), 
and vulnerability exploitation techniques (such as, \texttt{tcache} poisoning~\cite{silvio2019linux}, double free errors~\cite{doubly_owasp}, or null-byte overflows~\cite{2017null}).

\begin{figure}
    \centering
    \highlightbox{
    \textbf{Challenge Name:} ``zero\_to\_hero'' \hfill \textbf{Points:} 500 / 34201\\
    \textbf{Description:} ``Now you're really cooking. Can you pwn this service?.''\\
    \textbf{Goal:} Participants are given access to a binary and two libaries: \texttt{libc.so.6} and \texttt{ld-2.29.so}. To complete the challenge they must exploit the binary.\\
    \textbf{Solution:} To solve this problem, a participant must bypass the 2.29 \texttt{glibc} patch which fixes a double free vulnerability. To achieve this, they need to employ a null byte overflow to change a chunk's size, use this as a double free, and then conduct a \texttt{tcache} poisoning attack to overwrite \texttt{\_\_free\_hook} and succeed in the exploit.}
    \label{fig:pico_challenge}
    \caption{A sample security challenge from the picoCTF 2019 competition and a summarized description on how to solve it.}
\end{figure}

%% file: 4-methodology.tex
\section{Methodology}
Our data analysis was based on five focus groups conducted with each of the five top-scoring CTF teams from the picoCTF 2019 competition. 
We invited all members (a total of 25 individuals) to participate in the focus groups; 17 members participated. 
We followed a semi-structured interview format, and each focus group was scheduled for 1 hour and 30 minutes. 
Two researchers were part of each focus group: one researcher asked questions, while the other researcher took notes. 
Our questions were organized across eight themes, as described in ~\ref{methodology:script}. 
The focus groups were held online and the meetings were recorded and then transcribed. 
The transcriptions were then coded under a grounded theory approach~\cite{charmaz2014constructing}, following open and axial procedures by three independent coders, as described in \ref{methodology:coding}. 
Not all teams had full representation in the focus groups, which we discuss in \ref{discussion:limitations}, among other limitations.
\subsection{Focus Group Script}
\label{methodology:script}
Our goal was to uncover factors (e.g., practices, technologies, strategies) used by the top-ranking teams to achieve a successful collaboration. Due to the scarce literature, our interview script was built for breadth: covering a wide range of possible topics. Within these topics, we explored in more depth a particular topic if it seemed to be more salient (e.g., the team recurrently mentioned it) or if the team explicitly emphasized its importance. 
The list of topics we explored are as follows:
\begin{itemize}
    \item \textbf{Formation}: composition, recruitment, assembly mechanisms, as discussed by Harris et al.~\cite{harris2019joining}.
    \item \textbf{Technology}: team's use of current tools, their purpose, and desired/missing functionality.
    \item \textbf{Communication}: the media and purpose of members' communications.
    \item \textbf{Collaboration}: how, when, and why team members chose to work together.
    \item \textbf{Dynamics}: roles, responsibilities, and decision processes (e.g., leadership).
    \item \textbf{Coordination}: task definition, creation, assignment, and review and feedback processes.
    \item \textbf{Motivation}: which incentives motivated individuals to join and stay in their teams.
    \item \textbf{Limitations}: what limiting factors prevented teams/individuals from succeeding.
\end{itemize}

\subsection{Data Analysis}
\label{methodology:coding}
Our analysis was based on five focus groups (one per team) with a total of 17 participants. Three independent coders first listened to all focus group recordings and read each transcript. The lead author (who also led the focus groups) then grouped question/answer pairs across the 8 topics used in the interview script, keeping follow-up questions and answers together as appropriate.
Transcriptions were then coded by following an open coding approach to refine and reorganize our initial themes~\cite{charmaz2014constructing}. We allowed these codes to deviate from the original question categories, creating new categories as appropriate. Our approach follows Strauss' view ~\cite{corbin_basics_2007} of grounded analysis where researchers can start from an initial coding frame and continue to refine the code book as the they proceed.
After a first round of coding, the researchers compared codes and discussed similarities and discrepancies across their codes and interpretations.
The codes were then refined based on the discussions. We did not calculate inter-coder reliability as the goal was to identify emergent themes~\cite{mcdonald2019reliability}. Following, the coders jointly worked to cluster categories hierarchically (using axial coding), seeking to more narrowly categorize and sub-categorize codes. Through this process, we focused our discussion on action strategies and consequences. For example, why did a team make a certain decision, does this decision affect communication or dynamics? Additionally, we used the subcategories to enable the creation of a comparison table as a mechanism to laterally compare teams' practices and strategies. A result of this analysis can be observed in Table~\ref{tab:unmatched-estimation}. Lastly, we discussed teams' similarities and differences across each category. We describe our observations in the upcoming results section below.

\subsection{Ethics}
Our study was reviewed and approved by the Institutional Review Board (IRB) of our institution before any research activities began. 
As our research involved minors and voice recordings, we contacted the parents/guardians of the participants to obtain their written consent in addition to that of each participant. 
Prior to beginning the interviews, participants were once again briefed on the study, the data collection and retention policies, and ways for withdrawing from the study. 
Participation was voluntary and no financial incentives were provided for participation.

%% file: 5-results.tex

\begin{table}[h]
\resizebox{\columnwidth}{!}{
\centering
\begin{tabular}{clccl}
\toprule
\textbf{Placement} & \textbf{\begin{tabular}[c]{@{}c@{}}Location \& \\ Collaboration\end{tabular}} & \textbf{\begin{tabular}[c]{@{}c@{}}Prior \\ Collaboration\end{tabular}} & 
\textbf{Leadership} & \textbf{\begin{tabular}[c]{@{}c@{}}Task \\ Distribution\end{tabular}} \\ \midrule
1\textsuperscript{st}             & \begin{tabular}[l]{@{}l@{}}Distributed: \\ fully remote\end{tabular}          & Some                                                                    & Informal          & \begin{tabular}[l]{@{}l@{}}All specialists\\ with sub-specialists\end{tabular}  \\
2\textsuperscript{nd}                & \begin{tabular}[l]{@{}l@{}}Distributed: \\ fully remote\end{tabular}          & None                                                                    & Shared                                                         & \begin{tabular}[l]{@{}l@{}}All specialists \\ with sub-specialists\end{tabular} \\
3\textsuperscript{rd}                  & \begin{tabular}[l]{@{}l@{}}Distributed: \\ fully remote\end{tabular}          & None                                                                    & Shared                                                          & \begin{tabular}[l]{@{}l@{}}All specialists \&\\ basic generalists\end{tabular}  \\
4\textsuperscript{th}                 &
\begin{tabular}[l]{@{}l@{}}Mixed: \\ mostly remote\end{tabular}               & Some                                                                    & Hierarchical                                                               & \begin{tabular}[l]{@{}l@{}}All specialists\\ and one generalist\end{tabular}    \\
5\textsuperscript{th}                  & \begin{tabular}[l]{@{}l@{}}Mixed: \\ mostly remote\end{tabular}               & Some                                                                    & Informal                                                           & \begin{tabular}[l]{@{}l@{}}Mix of specialists\\ and generalists\end{tabular} \\ 
\bottomrule
\end{tabular}
}
\caption{
\label{tab:unmatched-estimation}
Performance and collaborative features of the top five ranking teams}
\end{table}

\section{Results}
\label{results}
We present the results from our analysis framed in the context of our research questions. With regards to performance (\textbf{RQ5}), we observed that many factors seemed to contribute to performance differently as described by the teams. As such, we discuss performance across each subsection, where pertinent. Commonalities across task coordination, dynamics, and collaboration, seem to be among the most influential factors in performance. 
Formation, motivation, and limitations were more heterogeneous across teams and individuals, but seemed to have a varying impacts on performance across individuals and teams. We assessed whether a certain factor had positive/negative impact on performance based on participants explicit descriptions (e.g., ``this strategy helped us work faster'') and through observation of implicit themes, such as teams describing how they spent too much time looking for information on their chat logs.

\subsection{Roles, Task Distribution, and Modularity}
Clearly defined roles based on specialties were the cornerstone of each team. Specialties were divided based on broad categories of problems (e.g., a reverse engineering specialist), which addresses our \textbf{RQ3}. 
Formation and recruitment centered around members' specialties. 
Because teams adopted a role-based task distribution approach, they recruited members based on specialty to maximize the coverage of tasks; these observations also shed light on our \textbf{RQ1}. 
Each member took ownership of a category of problems based on their specialty. 
This approach had multiple benefits. 
First, it increased efficiency by reducing redundant work (i.e., two members working on the same problem). 
Doing so, members were able to tackle more problems simultaneously.

\begin{displayquote}
T1--\textit{``When we were organizing the team each person kind of has a specific category they're going to do. So by keeping the work to each category, we kind of just eliminated the problem of double work.''}
\end{displayquote}

Second, each specialist was able to lead their section and determine how and when they needed help from other team members. 
This is important, participants said, because problems across categories may need different types or levels of involvement from other members that may be hard to define by a non-specialist. 
For example, a member working on a cryptography problem may benefit most from a teammate finding information on cryptosystems that might relate to the problem at hand. 
Whereas a member looking for a web security vulnerability might benefit more from somebody actively looking for potential vulnerabilities in the code. 
Third, it created a sense of accountability to the team. 
Members described not wanting to fall behind and drag their team down. 
Fourth, it allowed teams to be more modular, and thus if a member was unavailable for a competition, a similar specialist could more easily substitute for the missing member.

\begin{displayquote}
T1--\textit{``We have a poll for who wants to participate in the CTF, and then we try to organize teams based on different categories. So each person is better at certain category, and then we make sure there's one of each category on the team so they have everything covered.''}
\end{displayquote}

Fifth, it facilitated training of new recruits. A specialist would become the go-to expert for a category, and thus new recruits who are interested in that particular specialty can be trained by the more senior members.

While the specialist model carries several benefits, two scenarios may disrupt teams. 
If the competition or problems involve more categories than team members, members may have to take ownership of more than one category, which may become a heavy burden depending on the quantity and difficulty of the problems across the categories they own. 
Furthermore, if a specialist gets stuck in their category and no one else has any expertise in the area, progress becomes stunted because other members may not be able to offer enough help or the necessary help. 
A few teams mitigated these downsides by having members sub-specializing in other categories and/or by each member developing a general understanding of all categories (i.e., basic generalists). 
That way, to mitigate the first scenario, the team can assign two sub-specialists to a category. 
When they are done with their main commitments, they can direct their attention to their secondary category.

\begin{displayquote}
T4--\textit{If there's a five member limit, I always want two reversers slash binary exploitators on [the team], and then two web slash forensics people on [the team]. And then the last one is just everything else. Because if you have one person and one of those bigger categories, like web or reversing, it just gets troublesome, and they might get overwhelmed with the amount of stuff they have to do.}
\end{displayquote} 

In the second scenario (i.e., a member getting stuck), the specialist would first collaborate with the sub-specialist. 
If they are unable to solve the problem together, they would then involve the other members. 
Since the other members are generalists, then they can offer additional insights and also know how to better support (e.g., whether to write code, or what online articles may be related). 
Thus, while having highly talented individuals in a team is important, having the right team composition and task distribution makes the team significantly more efficient.

\begin{displayquote}
T2--\textit{``We didn't really have anyone that specialized in reversal problems. We all worked together on that. We all bounced ideas around and checked. Because we do have an overlap in skills. That helps a lot, if we're stuck.''}
\end{displayquote} 

\textbf{\textit{Observation \#1:}} Teams organize and distribute tasks across roles based on computer security specialties (e.g., web security, cryptography) to reduce redundant work, increase team modularity, distribute accountability, and to allow each member to define how they want their teammates to provide help when needed.

\subsection{Performance, Trust, and Leadership}
Performance in previous competitions was the most important signal that members use to determine the credibility of potential recruits. Within the team, this indicator may also be used implicitly to determine leadership and as a proxy for trust as defined in Section~\ref{sec:related-work} (i.e., ``expectation that others will follow through with their commitment''). These results shed more light on our \textbf{RQ3}.

\begin{displayquote}
T5--\textit{``The best way to see how someone would do on a CTF is to just look at their performance on previous ones.''}
\end{displayquote}
\begin{displayquote}
T1--\textit{``Competing in past CTFs before would be a really good indicator, I think. And of course, specialization in a category that we need more skill in for the team.''}
\end{displayquote}

This signal is easily quantified by three factors: the number of problems an individual has solved, the difficulty of the problems solved, and the speed to solve these problems. 
This information is typically available at the team level. 
However, it also informally transmitted by word-of-mouth or inferred by an individual's participation in previous CTFs or their contribution of write-ups (tutorials on how to solve specific problems). 
This metric, because it is seen as an objective measure of individual abilities, is used when recruiting new members as a predictor of their performance. 
Additionally, it seems as if performance can be used as a mediator for trust between members. 
That is, because members take ownership of task portions, other members may feel more at ease if they perceive their teammate(s) as more capable, based on their past performance. 
Similarly, we observed that informal leadership across teams seemed to gravitate to members who had higher performance signals. 
On the other hand, we also observed that leadership was more distributed across teams were all members seemed to have similar performance signals. For instance, members of T1 mentioned one teammate as being the ``unspoken leader.'' This member also had substantial experience participating in other competitions and doing bug bounties. Similarly, T5 also had two members who did not denominate themselves as leaders, but made a lot of the decision in terms of organization and task distribution. T4 had a similar member as T1, but in their case they decided to denominate this member the team captain. In the case of T2 and T3, all members rated their abilities at the same level. At the same time, both of these teams described fully democratic decision-making processes: ``someone has an idea and we all agree or we disagree.''

\textbf{\textit{Observation \#2:}} Past CTF performance is used as a strong determinant of future performance and thus, is an important metric used during member recruitment, along with role. 
An individual's past performance seemingly correlates with their perceived capability. Consequently, members with higher perceived performance or knowledge, may informally raise as leaders unless the perceived performance/abilities is balanced across members.

\subsection{Technology Usage and Needs}
Apart from traditional voice and text-based communication needs, all teams mentioned the need for a mechanism to keep track of task ownership and progress, which addresses our \textbf{RQ2}. 
All teams used Discord as their primary communication platform. 
Discord is a communication software which features text and VoIP channels, with rich integration capabilities and fine-grained access control.\footnote{https://discord.com/} 
Apart from text and voice, users can share images, files, and stream their screen with other members of the server. 
Teams used text channels for asynchronous communications and disjointed conversations on various topics, with the option to directly talk about something using the voice channel.

Task visibility and tracking was a harder problem, and teams experimented with different approaches. 
One team used Trello, a Kanban-style web-based application where users can make lists, assign roles, and specify deadlines~\footnote{https://trello.com}. 
The team mainly used it to leave notes under each problem, but seemed to not find it very valuable.

\begin{displayquote}
T3--\textit{``[Trello] was okay I think. We didn't use Trello for [a subsequent CTF]. So I don't think it was completely necessary [for picoCTF]. But it was sort of helpful to keep track of which problems people are working on and also for keeping track of notes so we don't have to keep searching through the Discord history.''}
\end{displayquote}

Another team used Google Sheets for the same purposes. 
The other teams mostly relied on Discord, but often griped about conversations being mixed up, or retrieving information from the chat history. 
Alternative strategies involved creating a channel for each problem (the team mentioned it did not work well) and creating a channel for each category, which had better results, but was still prone to information retrieval issues. 
Furthermore, teams also mentioned they lacked a good way of sharing larger files or code snippets.

\begin{displayquote}
T5--\textit{``Okay, so one annoying thing about Discord [for sharing code] is that there's a message limit of 2000 characters... it'll just complain and say you have to upload a file instead. [Another problem is that] I don't think any other platform even has like code highlighting, you can't even put code blocks with highlighting although maybe Slack can do that?''}
\end{displayquote}

\textbf{\textit{Observation \#3:}} Text and voice communication were adequately addressed by existing software. On the other hand, teams struggled with task visibility (i.e., awareness of what each member is working on), tracking (i.e., tracking progress and resources on each task), and resource sharing and retrieval (e.g., code and files). Teams experimented with various software but no tool seemed to suit all their needs.


\subsection{Formation, Motivations, and Limitations}

Assembly mechanisms, reasons to join together, and familiarity between members were different across all teams: from friends who went to the same school, to strangers who found each other through CTF-oriented Discord channels (\textbf{RQ1}). 
On the other hand, motivations (why members sought to do well) across all teams and members were mostly similar (although ranked differently by each member), which addresses our \textbf{RQ4}. 
Members often cited a combination of incentives: educational or professional development, entertainment, socialization, prizes, and measuring performance. 
These factors (including assembly) seemed to have little impact on performance and collaboration as expressed by the teams, except for one team (T3). 
One team mentioned that they formed the team and recruited members with the goal of earning a top score in the competition. 
When screening recruits, the founding members as well as the prospective recruits explicitly stated that they wanted to ``place'' in the competition (i.e., enter the top ranking in the score ladder). 
This common motivation, as described by participants, played a role when preparing and during the competition. 
\begin{displayquote}
T3--\textit{I had been growing my skills from last year. And so I thought if I joined a team, I'd have, like they said, a chance of getting up there in the top five. And so when I saw that message reaching out from a strong team, it sounded like a good opportunity.}
\end{displayquote}

\begin{displayquote}
T3--\textit{``Yeah, for me, it was mostly the same reasons. I wanted to place and I also saw that they were running [their own CTF]. It showed that they were good. So that also influenced me. I think it has to do with sort of the goal. We were trying to place really well, as a competitive team. ''}
\end{displayquote}

Setting the common goal of doing well in the competition seemingly increased their motivation, and potentially their performance as well, as it increased the accountability of each member in achieving that goal. 
Furthermore, since the team was fully democratic, we hypothesize that having an aligned goal facilitated decision-making since the interests of one member would not conflict with another.

\textbf{\textit{Observation \#4:}} While motivation across team members was heterogeneous across most teams, it reportedly played a big role in the performance of one team (T3). This suggests that while common motivation might not be a requirement for a successful collaboration, it may become an enhancer.

While most discussions on limitations were straightforward (e.g., lack of technology functionality, lack of expertise/time for certain problems), some members discussed their gripes with diversity and their social identity self-presentation, which adds another dimension to our \textbf{RQ4}. 
CTF competitions, because they are conducted online and communication in larger groups is typically text-based, often allow participants to limit their self-presentation to a mere online handle. 
Members from two teams mentioned that this anonymity was relieving, since their abilities were not subject to presumptions based on things like their gender or race. 
However, when this anonymity was no longer in place, they felt pressured to compensate based on their social identity. 
Across the five teams (25 participants), only three participants were female (two of which were present in the focus groups) and only one participant was from an under-represented group.

\begin{displayquote}
P0\footnote{We omit mentioning the team of the above two members (P0 and P1) to preserve their privacy.}--\textit{``I guess being aware that I'm the only female in a group, might add a bit of pressure I suppose. Like sometimes I think that if I say something strange or if I mess up somewhere, I might be confirming the assumption that females aren't as good at CS, or maybe I might be confirming certain stereotypes about my gender.''}
\end{displayquote}

\begin{displayquote}
P1--\textit{``I think the online aspect where we're just all like, you know, playing together, nobody knows where you're from definitely helps or at least makes it ([issues of diversity]) less of a problem.''}
\end{displayquote}

\textbf{\textit{Observation \#5:}} Members may self-censor their self-presentation (particularly their social identity) across online platforms to avoid presumptions of their abilities.

\textbf{\textit{Observation \#6:}} Members may feel pressured to perform to a certain level to prove that their social identity does not inhibit their abilities.

%% file: 6-discussion.tex
\section{Discussion, Recommendations, and Future Work}
The security teams in this study developed a variety of strategies to cope with the multiple known challenges related to dispersed collaboration. 
At the core of these strategies was the usage of roles to determine coordination (\textbf{Obs.~1}), communication (\textbf{Obs.~1}), recruitment (\textbf{Obs.~2}), and technology usage (\textbf{Obs.~3}). 
This role-based approach attenuated many of the known downsides of dispersed teams, such as by reducing interdependencies and the amount of coordination and communication needed. 
We also found that teams assembled in a variety of ways, through various channels, and for various reasons (\textbf{Obs.~4}), but that quantified performance and roles heavily affected the assembly process (\textbf{Obs.~2}). 
Finally, we found that participants from under-represented groups faced a set of unique challenges that ought to be accounted for in the design of online communities (\textbf{Obs.~5-6}). 
Our observations shed light on all of our proposed research questions (\textbf{RQ1-5}). 
We next discuss the implications of our results across various aspects, make suggestions, and identify areas which merit future exploration.

\textbf{Security and Software Engineering Teams.} 
While many of the problems caused by dispersion documented in the context of software engineering might readily explain the issues that security engineering teams may face, it does not seem like the same remedies extend.
We find that security teams cope with dispersion challenges differently than their software engineering counterparts. 
In our study, teams employed a role-based approach based on separate specialties to determine to determine work distribution and collaboration interactions (\textbf{Obs.~1}).
Such an approach is likely not possible for software engineers without a prior planning due to the interdependencies that arise when building software and the lack of clear cut roles.

Thus, because the nature of the work and its requirements are different, treatments conceptualized in the context of software engineering teams (such as radical collocation~\cite{teasley2002rapid}) may have a lesser impact on security teams. 
In fact, two teams in our panel had the possibility to work in the same physical space, yet still opted for asynchronous remote work.

We believe that team-oriented research for security teams is imperative given the continuous growing need for security services, professionals, and teams. While our study focused on teams that encompasses various capabilities, we note that there exist various other types of security teams which remain unexplored, such as red/blue teams and penetration testers, both in industry as well as in crowd work contexts.

\textbf{Technology Needs.} Different practices tend to imply different needs, and consequently call for different supporting tools. 
Given the lack of academic work on security teams, it is unsurprising that there are no tailored tools for the collaboration needs these teams described.
This is evidenced in our study, whereby teams juggled a variety of tools to cope with their task visibility and knowledge sharing needs (\textbf{Obs. 3}). In software engineering, real-time shared environments have been a popular approach to improving visibility and mitigating coordination costs~\cite{lee2017exploring,salinger2010saros,latoza2014microtask}.
For security teams which mostly focus on reviewing code, existing issue tracking systems (such as GitHub's) might be adequate if the project allows for a priori task segmentation.
However, for faster paced environments, such as a newly launched bug-bounty program, non-real-time issue tracking systems may fall short. Because there may not be sufficient information (e.g., which parts of the codebase contain cryptography bugs) nor time for a priori coordination, short-term security teams would likely benefit from real-time systems which provide information on the portions (e.g., code segments) that other analysts might be working on and the types of tasks or bugs they are testing for (e.g., web bugs). Additionally, it is unclear what tools readily and acceptably extend to other tasks beyond code reviews, such as testing remote server deployments. How should tests of applications or configurations be tracked and made visible? Or how should analysts share helpful resources and strategies?

\textbf{Teams in Security Crowd Work.} 
In the context of security crowd work, and in particularly bug-bounty programs, the current paradigm is focused on harnessing the expertise of various individuals. 
However, the current system is inefficient, and invites a lot of duplicate and invalid reports caused by competition among individuals~\cite{zhao2016crowdsourced}. 
Our study cases that short-term security teams are viable and could help mitigate efficiency issues, as evidenced by team strategies to reduce duplicate and redundant work. Participants in our study were able to efficiently self-organize relying only on roles and performance (\textbf{Obs. 2}), two signals that are already available across bug bounty platforms.
Allowing users in bug-bounty programs to self-assemble may allow for a more efficient allocation of users' abilities within a program.
Thus, rather than all participants analyzing everything at the same time, crowd workers can spend more dedicated effort in specific sections. 
Furthermore, the effectiveness of the program is also poised to increase, as crowd workers can now collaborate and share knowledge, potentially increasing the complexity of bugs they are able to catch while also speeding up the process. 
Finally, we also envision a benefit across crowdsourced security communities, as team assembly might motivate and allow for more opportunities for professional development and education of newer crowd workers, which have been motivating factors both for our participants (\textbf{Obs. 4}), as well as for joining BBPs according to bug-bounty hunters~\cite{akgul2020hackers}.
Naturally, questions of resource pooling or team-level competition may arise in this new paradigm. To further explore this proposition, a field study with bug-bounty hunters in teams should be conducted.

\textbf{Diversity and Self-Censoring.} Women and participants from under-represented groups in our study saw the anonymity shield granted by online participation as a beneficial feature. 
This anonymity allowed them to participate in online CTF communities without having their gender or race being subject to the judgements or presumptions that other members in the community may have. 
To achieve this, however, members may self-censor their presentation (\textbf{Obs. 5}) across these communities or even within \emph{ad-hoc} teams they become part of. 
Within teams, they may also feel pressured to perform to a certain standard (\textbf{Obs. 6})---which is basically relatively close to being an impostor syndrome of sorts. With regards to majority members in the CTF community, perhaps there is also an issue of abstaining from diversity such as the one described by G\'{o}mez-Z\'{a}ra et al.~\cite{gomez2020impact}, which under-represented participants may be experiencing.
Past research has shown that brief social-belonging interventions can improve academic and health outcomes of under-represented students~\cite{walton2011brief}. These observations could raise questions on how to design online communities, particularly in specialized crowd work settings, to mitigate feelings of exclusion based on social identity. For example, sites like HackerOne or Bugcrowd may consider always keeping personal profiles optional. We recognize that our sample is already small and under-represented  participants were also a minority in our sample (3 out of 17 participants). While we do not claim that this is a phenomenon which affects all CTF players from under-represented groups, a previous experiential paper on using CTF problems with high school students found a significant difference between male and female students' experiences. The researchers note that ``the male students thought the [CTF] activities were more enjoyable and interesting than female students''~\cite{jin2018game}. Given our results, we believe that this issue is not due to lack of interest, but rather related to the pressure to perform our participants describe to this issue. Future studies ought to explore this issue to begin addressing issues of diversity in the education and profession of computer security.

\subsection{Limitations}
\label{discussion:limitations}
We identify two main limitations with regards to our study and the implications of our findings. 
The first is that the teams we observed (teams formed for a CTF competition) are not a strict subset of neither industry security engineering teams nor of teams of bug-bounty hunters, and as such, may affect the generalizability of our results. 
Because bug-bounty teams remain, at the time of writing, 
quite rare compared to solo operations, 
we argue that CTF teams are a good proxy to study the feasibility of implementing team-based bug-bounty programs.
We posit that CTF teams are a better proxy than industry computer security teams because the CTF teams we studied mostly self-assembled online, had no prior collaboration, and worked on a time-constrained task (two weeks). 
Additionally, bug-bounty hunters are often CTF players as well. For instance, members across the teams we studied had submitted more than 50 bug reports at the time of interview for organizations such as the Department of Defense. Furthermore, CTF challenges are increasingly being used to train practictioners in computer security across classrooms~\cite{mirkovic2014class} and even by bug-bounty programs on their educational platforms~\cite{hacker1012020,levelup}.

Secondly, not all members were able to participate in the focus groups. 
In particular, one team was only represented by one member. 
We tried to account for this by framing the focus group around collective group practices, i.e., practices that all members took part of so that each member was able to talk about them. 
For example, a single member was still fully capable answering questions about the technology needs of their team, their self-assembly, their communication practices, etc. 
While we believe this approach allowed us to have a good picture of each team, we note that in some cases having additional responses from each member added nuance to the team dynamics.

%% file: 7-conclusion.tex
\section{Conclusion}
We presented a first view on the practices and needs of the security teams who participated in an online security competition (picoCTF 2019). 
The teams we studied organically developed many organizational algorithms to maximize their efficiency and reach a top spot among a very large number of competitors. 
Surprisingly to us, the top three teams seemed to be unhindered by the many challenges posed by dispersed collaboration. 
Equally surprising is that the fourth and fifth place teams, despite being in the same geographic area (students at the same school), opted to conduct most of their work online, collaborating asynchronously. 
These teams placed member specialization at the core of their decision-making, from formation to task distribution. 
This approach seemed to yield performance gains in various aspects, such as the ability to get unstuck from areas in which the team lacked expertise, or the ability to minimize redundant work. 
Beyond providing a first glance at the practices and needs of successful security teams, our findings shed light on a possible path forward for the future of the emerging field of crowdsourced security. 
Scaling crowdsourced security offerings beyond individuals not only has the potential to increase the efficiency of tasks, but can likely offer many benefits to the overall community. For instance, more opportunities for crowdworkers' development may arise, through knowledge sharing. 
At the same time, some of our findings---such as the poignant examples of members preferring not to disclose their gender lest their competences were called into question---make resoundingly clear the importance to assess issues of exclusion based on social identity, particularly in collaborative crowd work environments.
Our work also sheds light on the large research gaps that exist in studying security teams: the area is ripe with opportunities for team-oriented scholarship and technology development. 
From the teams' technology usage---adapting and reworking tools which just do not seem to fit the bill---to better understanding to which extent our findings carry over to professional bug-hunters, to corporate security teams, there seems to be plenty of room for improvement, particularly as we continue to learn more about the practices of more types of security teams.